\documentclass[useAMS,usenatbib,a4]{mn2e}
\usepackage{graphicx}
\usepackage[usenames]{color}
\usepackage{epstopdf}
\title{M dwarfs and the fraction of high carbon-to-oxygen stars in the solar neighbourhood}

\author[J. E. Gizis, Z. Marks, P. H. Hauschildt]{John E. Gizis$^{1}$\thanks{E-mail:
gizis@udel.edu (JEG)}, Zachary Marks$^{1}$, Peter H. Hauschildt$^{2}$\\
$^{1}$Department of Physics and Astronomy, University of Delaware, Newark, DE 19716, USA\\
$^{2}$Hamburger Sternwarte, Gojenbergsweg 112, 21029, Hamburg, Germany}

\begin{document}

\date{Submitted 9 July 2014, revised 18 September 2015}
\pagerange{\pageref{firstpage}--\pageref{lastpage}} \pubyear{2015}

\maketitle

\label{firstpage}

\begin{abstract}
We investigate the frequency of high carbon-to-oxygen (C/O $= 0.9$) M dwarf stars in the solar neighbourhood. Using synthetic spectra, we find that such M dwarfs would have weaker TiO bands relative to hydride features. Similar weakening has already been detected in M-subdwarf (sdM) stars. By comparing to existing spectroscopic surveys of nearby stars, we show that less than one percent of nearby stars have high carbon-to-oxygen ratios.  This limit does not include stars with C/O$=0.9$, [m/H]$>0.3$, and [C/Fe]$>0.1$, which we predict to have low-resolution optical spectra similar to solar metallicity M dwarfs. 
\end{abstract}



\section{Introduction}

The carbon-to-oxygen ratio plays a key role in chemistry at low temperatures. In the Sun, there are approximately two oxygen atoms for every carbon atom \citep{2005EAS....17...21G,Asplund:2009gf,2010A&A...514A..92C}, but the measurement of carbon and oxygen abundances remains difficult in the Sun and in other stars. There has been considerable interest in the existence of high carbon-to-oxygen ratio planets \citep{Kuchner:2005kx,Madhusudhan:2012yq,2013ApJ...763...25M}, which would form more easily if carbon is enhanced relative to oxygen in the original nebula, and hence host star.  Even if carbon atoms do not outnumber oxygen atoms, the mineralogy of planets formed from C/O$ > 0.8$ gas will be distinct from siliicate-dominated planets \citep{2010ApJ...715.1050B}.  

Some recent determinations of the carbon-to-oxygen ratio in G dwarfs find that $\sim 25\%$ of nearby stars have C/O $> 0.8$ \citep{2010ApJ...715.1050B, Delgado-Mena:2010yq,Petigura:2011zr} and $\sim 8\%$ have C/O $>1$, although it should be noted that some of these have C/O$=0.65$ for the Sun rather than the more recent C/O$=0.55$, and therefore those reported C/O ratios are shifted towards higher values. \citet{Fortney:2012lr} has suggested that the frequency of carbon-rich solar-type stars in these studies are over-estimated, pointing to the difficulty of measuring oxygen and the low observed frequency of dwarf carbon stars. The latter is a significant constraint on the existence of low-mass stars with C/O$>1$. Most of the oxygen in the atmospheres of cool ($T_{\rm{eff}} < 4000$K) stars would be locked up in CO, and the emergent spectra would be dominated by carbon molecules instead of the usual oxygen molecules (TiO, H$_2$O).  As a result, they would not be classified as M dwarfs even in low-resolution optical spectra.  \citet{Fortney:2012lr} concludes that only 10\% of stars have C/O$>0.8$ and $\sim 1-5$\% have C/O$>1$, and cautions that this is likely an overestimate.   \citet{Nissen:2013fj} measures 33 G dwarfs using  high resolution spectra including a different oxygen line, and concludes that the claims of high C/O are ``spurious." On the other hand,\citet{2013ApJ...778..132T} present a case study of the 55 Cnc (G8V) planetary system and conclude that C/O $\approx 0.8$.  

In this study, we investigate the possibility that $\sim10\%$ of M dwarfs have C/O$>0.8$. In Section~2, we show that high carbon-to-oxygen ratios produce dramatic effects on synthetic spectra that would easily be detected at low spectral resolution.  
In Section~3, we use this result to show that a high C/O ratio is very rare ($<1\%$) amongst solar neighbourhood M dwarfs.   We discuss our results in Section~4.

\begin{figure*}
\includegraphics[width=\textwidth]{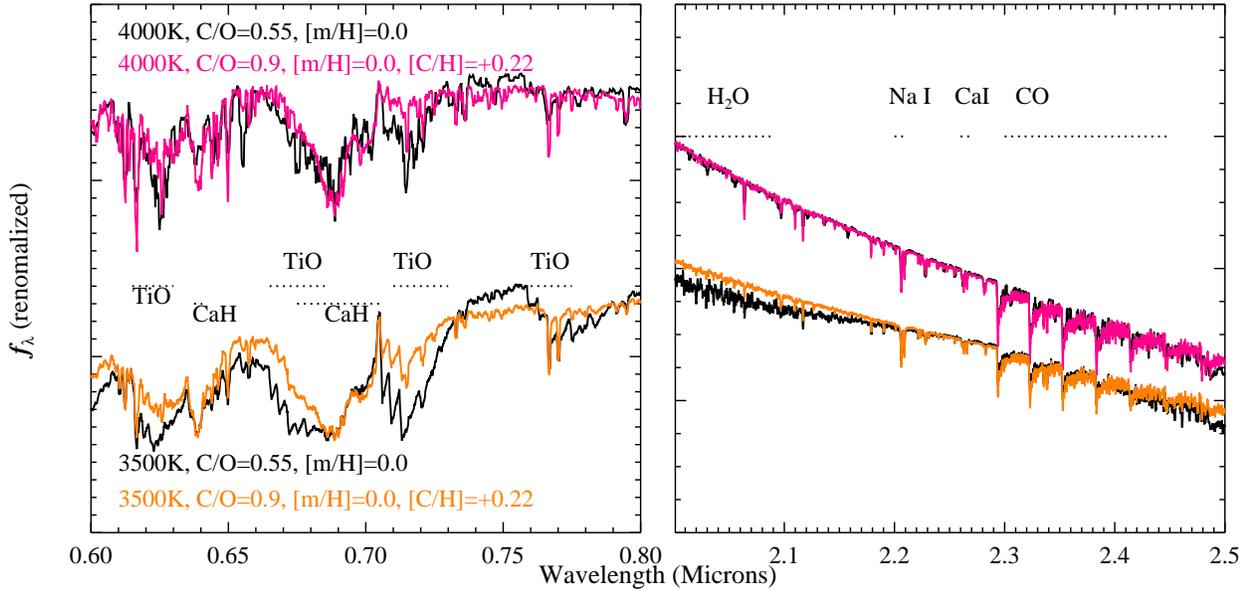}
\caption{Synthetic spectra of model solar metallicity M dwarfs (black) and enhanced carbon-to-oxygen model M dwarfs (colored) for both 4000K and 3500K. Left: Red optical, with major TiO and CaH bands are marked.  Right: Near-infrared, with H$_2$O, Na doublet, Ca triplet, and CO bandheads marked. \label{fig-synthetic1}}
\end{figure*}

\begin{figure*}
\includegraphics[width=\textwidth]{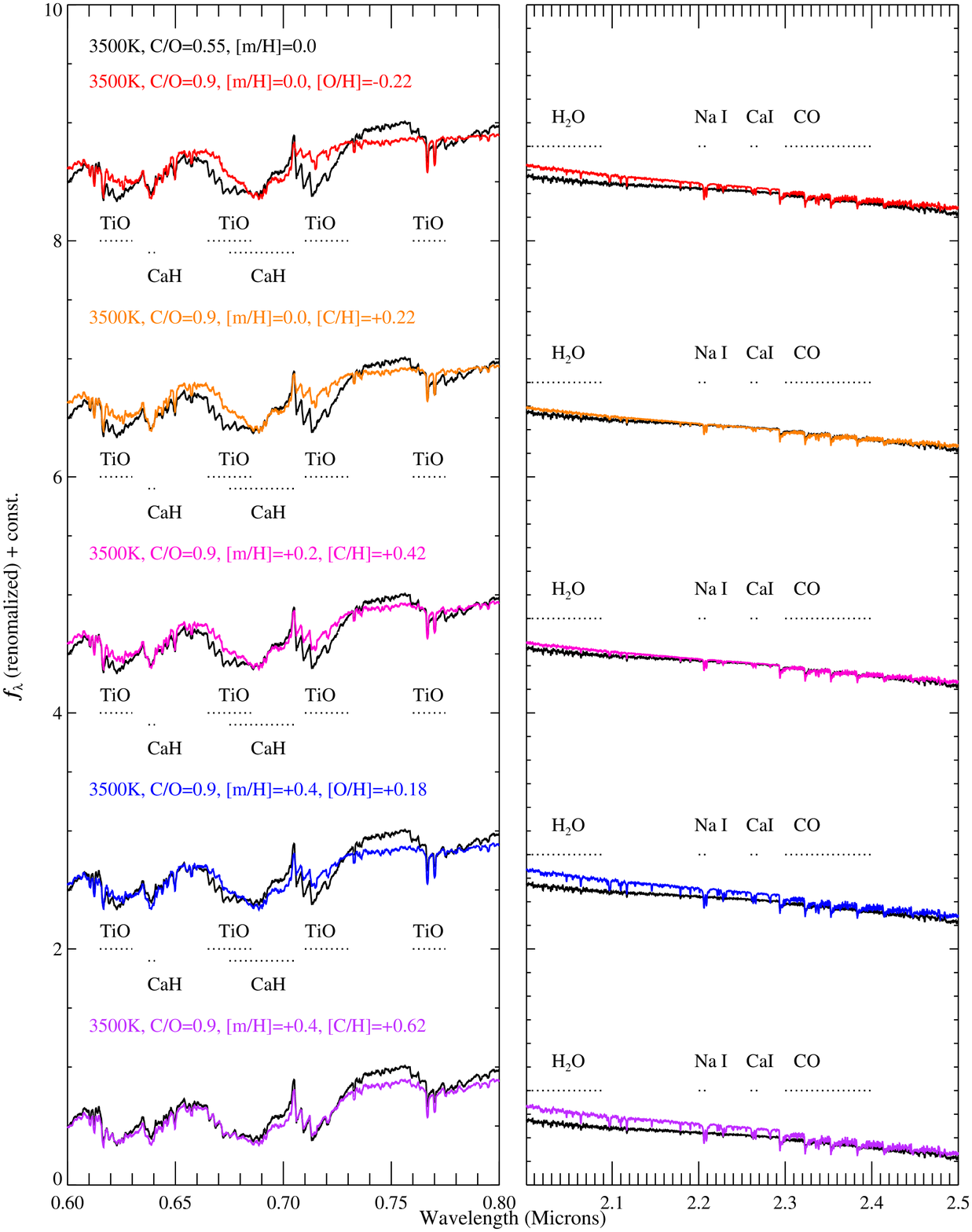}
\caption{Synthetic spectra of a model 3500K solar metallicity M dwarf (black) and five high carbon-to-oxygen model M dwarfs (colored). Left: Red optical, with major TiO and CaH bands are marked.  Right: Near-infrared, with H$_2$O, Na doublet, Ca triplet, and CO bandheads marked. The estimated $\zeta$ values (see text) from top to bottom 
are 0.3, 0.5, 0.6, 0.6 and 0.8.  
\label{fig-synthetic2}}
\end{figure*}

\section{Model Results\label{models}} 

M dwarf spectra are dominated by simple molecules, making them sensitive to carbon/oxygen chemistry.
Indeed, \citet{Allard:2012ul,2013MSAIS..24..128A} have shown that the new, lower solar oxygen abundance \citep{Asplund:2009gf} improves the agreement between synthetic and observed spectra of typical disk M dwarfs. \citet{2014PASJ...66...98T} discuss the measurement of [C/Fe] in M dwarfs.  We are interested in the possibility of a spread in the carbon-to-oxygen ratio in nearby disk stars, and particularly in the fraction of carbon-rich (C/O $\approx 0.9$) stars. Synthetic spectra were computed using the PHOENIX code, which has a long history of successful applications to M dwarf spectra \citep{1999ApJ...512..377H}. Our version is very close to the code used for the model library presented by \citet{2013A&A...553A...6H}.  The carbon-rich models have less oxygen or additional carbon such that the carbon-to-oxygen ratio is 0.9. 
Figure~\ref{fig-synthetic1} shows the results for 4000K and 3500K models.  
Because many of the G dwarfs with reported high C/O ratio also have higher than average overall metallicity (see the \citealt{2014AJ....148...54H} compilation), we also computed metal-rich models in which all metals are increased to [m/H] $= +0.2$ or [m/H]$ = +0.4$, and then additional carbon is added, or oxygen is removed, to make C/O$=0.9$.  High metallicity and high C/O G dwarfs typically have lower oxygen than other metals, so the [m/H]$=+0.4$, [O/H]$=0.18$ are more relevant than the [m/H]$=+0.4$, [C/H]$=0.62$ models. These 3500K models are shown in Figure~\ref{fig-synthetic2}. The models are internally consistent, with the model structure iterated to be consistent with the C/O ratio.  

Although real stars will have a variety of abundance anomalies, these simple models are sufficient to demonstrate the importance of enhanced carbon. As one would expect, the carbon-rich models have a stronger CO band head at 2.3 microns and weaker H$_2$O bands, because more of the oxygen can be locked up in CO by carbon. 
\citet{Rojas-Ayala:2010uq} showed that metal-rich, scaled solar abundance stars have enhanced Na and Ca atomic lines in the K-band, and we observe the same trend in our metal-rich models.  The effects in the red part of the spectrum are even more dramatic. TiO is much weakened while the depth of other molecular features, particularly those due to CaH, are practically unchanged.  Even in low-resolution spectra, the changes in the TiO bands at 0.66-0.68 $\mu$m and 0.71-0.72 $\mu$m should be very noticeable. The effect is strongest when oxygen is reduced, but even in the high metallicity $[m/H]=+0.2$ models with enhanced carbon the hydrides are considerably stronger relative to TiO. However, for the very highest metallicities ([m/H]$=+0.4$), the effect is strong only for the reduced oxygen ([O/H]$=+0.18$, [C/H]$ = +0.40$) model.  For [m/H]$=+0.4$ with enhanced carbon ([O/H]$=+0.4$, [C/H]$ = +0.62$), the changes are subtle and probably not obvious at low resolution, but this composition is probably not realistic. In the Hypatia catalog \citep{2014AJ....148...54H}, 48\% of the stars with C/O$>0.8$ have [Fe/H]$>0.3$ and [O/Fe]$<0$, but only 0.4\% of stars with C/O$>0.8$ have [Fe/H]$>0.3$ and [C/Fe]$>0.1$.

\begin{figure}
\includegraphics[width=0.5\textwidth]{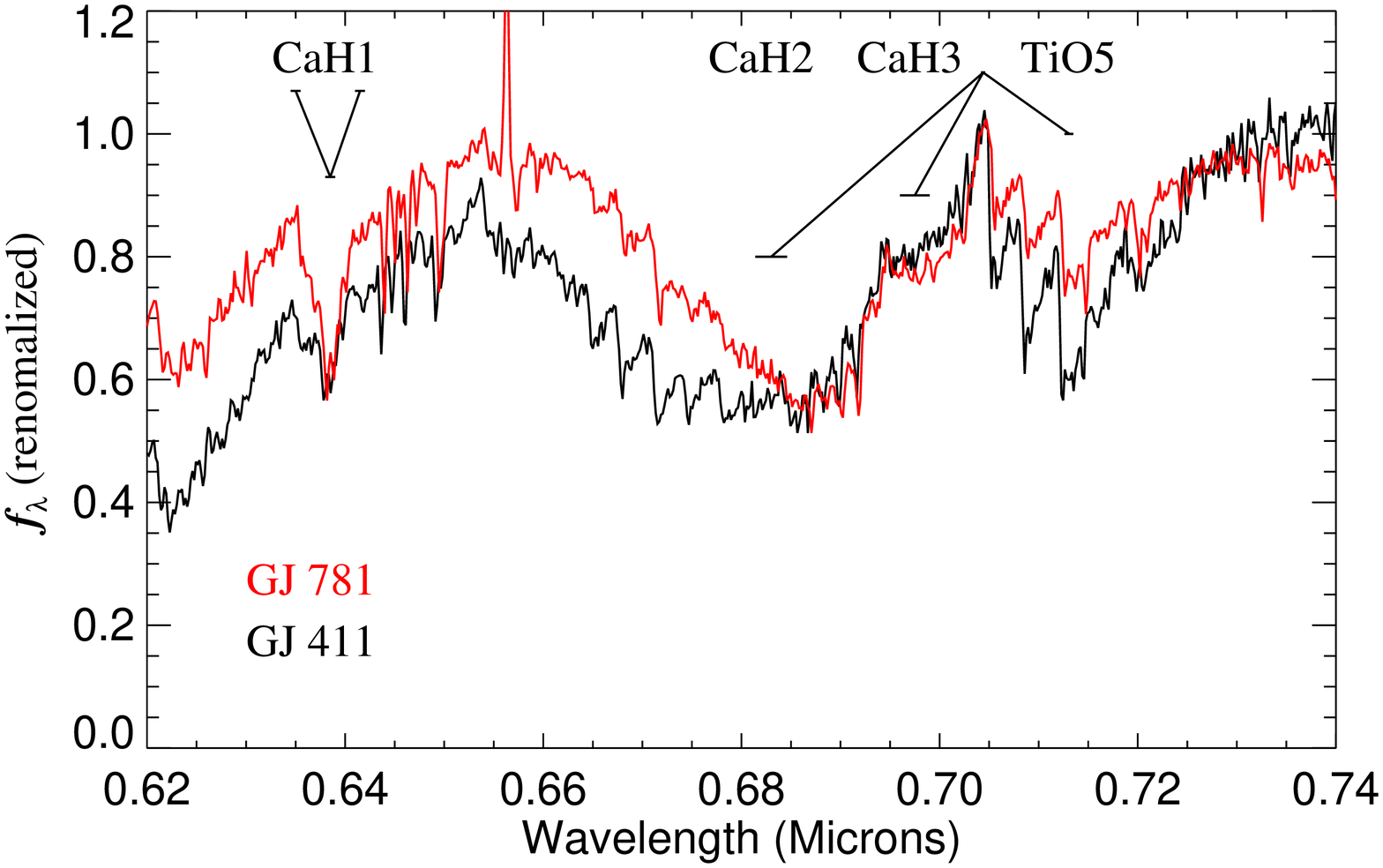}
\caption{Our comparison of the spectra of the M2 V standard GJ 411 and the sdM1.5 GJ 781. The wavelength regions used to measure TiO and CaH indices are marked. GJ 411 has TiO5$=0.60$, CaH1$=0.82$, CaH2$=0.53$; GJ 781 has TiO5$=0.77$, CaH1$=0.72$, CaH2$=0.56$. The original source of these spectra and indices are \citet{1995AJ....110.1838R}. \citet{2013AJ....145..102L} finds $\zeta = 0.58$ for GJ 781.  
The weak TiO bands are similar to the carbon-rich models in Figure~\ref{fig-synthetic2}, but the balance of evidence suggests that GJ 781 is a metal-poor subdwarf.  
\label{figreal}}
\end{figure}

Qualitatively, in the optical these differences from solar-metallicity are similar to those seen in metal-poor stars, particularly the sdM class \citep{1997AJ....113..806G} of M subdwarfs. Figure~\ref{figreal} compares a typical disk M dwarf, the M2 V standard {GJ 411}, with the sdM1.5  {GJ 781}. The TiO is weaker relative to the CaH, with an offset similar to that predicted for our carbon-rich models. We can therefore leverage the existing studies of the space density of optically-classified M dwarfs and M subdwarfs to constrain the fraction of M dwarfs that may be carbon rich. 

\section{Observational Constraints\label{spacedensity}}

Band indices measuring the TiO and CaH feature have been the basis of modern M subdwarf classification systems \citep{1997AJ....113..806G,2007ApJ...669.1235L} and have been used by many different groups, though \citet{2008AJ....136..840J} have criticized the method for also being sensitive to surface gravity. The TiO5 index measures the feature at 0.704-0.714 $\mu$m and the CaH2 index measures the $0.681 - 0.704$ $\mu$m feature \citep{1995AJ....110.1838R}, with smaller values indicating more absorption. The $\zeta$ parameter uses these indices to construct a metallicity indicator independent of temperature \citep{2007ApJ...669.1235L}. The carbon-rich models shown in Figure~\ref{fig-synthetic2} predicted a numerically larger TiO5 index at a given CaH2 (or $T_{\rm{eff}}$, which is less easily measured).  This would also be characteristic of a moderately metal-poor M subdwarf (the sdM class), and would appear as a small $\zeta$ parameter in the system.  We do not consider the models accurate enough to precisely compare $\zeta$ values, but we can use the observed frequency of M dwarfs/subdwarfs with weak TiO to set limits on the fraction of carbon-rich M dwarfs; that is, we assume that the qualitative model result is correct. We approximate $\zeta$ by computing the band indices for each model in Figure~\ref{fig-synthetic2}, adopting TiO5$_\odot$ from the solar metallicity model,  so that $\zeta = {{1-{\rm TiO5}}\over{1-{\rm TiO5}_\odot}}$.  All the models have approximately the same CaH2 and CaH3 indices, but TiO5 has larger values, leading to $\zeta < 0.7$ for each model, except the metal-rich, carbon-enhanced model which has $\zeta = 0.83$. The exact values should be viewed with caution, but the trend is clear.

\begin{figure}
\resizebox{0.88\linewidth}{6cm}{\includegraphics{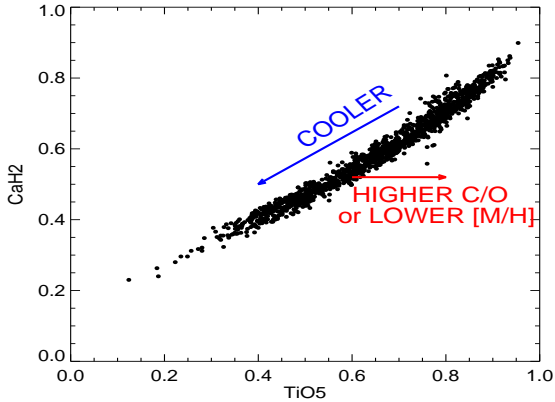}}
\caption{The observed strength of the TiO band at 0.704-0.714 $\mu$m (the TiO5 index) and CaH at 0.681 - 0.704 $\mu$m (the CaH2 index) for all known 1,405 bright ($J<9$) northern hemisphere M dwarfs with spectroscopy \citep{2013AJ....145..102L}. Smaller numbers indicate more absorption. The constant temperature models shown in Figure~\ref{fig-synthetic2} have $CaH2=0.5$ but are displaced horizontally with larger TiO5 index for high C/O. 
\label{figbands}}
\end{figure}

Most important for our purposes is the complete, well-defined spectroscopic survey of the 1,405 brightest ($J<9$) northern hemisphere M dwarfs \citep{2013AJ....145..102L}, which provides TiO5, CaH2, and $\zeta$ (Figure~\ref{figbands}).  The brightest sdM star in that sample is GJ 781.  GJ 781 has a high space velocity and lies in the subdwarf region of the H-R diagram \citep{1997AJ....113..806G}; it has an unseen white dwarf companion \citep{1998AJ....115.2053G} and a metal-poor T8 brown dwarf companion \citep{2013ApJ...777...36M}. Another small $\zeta$ parameter star is LSPM J2319+7900S. This star is a common proper motion companion to HD 220140 \citep{Gould:2004qf}, which is a mildly metal-poor G dwarf ($[Fe/H]= -0.52$, $[\alpha/H] = 0.20$, \citealt{Casagrande:2011ve}). Overall, 29 of 1399 objects have $\zeta \le 0.80$, suggesting that only $\sim$2\% of this sample could be high C/O ratio M dwarfs, but many (if not all) are metal-poor.  A more heterogenous, and partially overlapping, sample is the two thousand late-K and M dwarfs within 25 parsecs observed by the PMSU survey \citep{1995AJ....110.1838R,1996AJ....112.2799H}, which approximates a volume-limited sample. Of these stars, less than one percent show the sdM type (including Kapteyn's Star, {GJ 191}), and these are consistent with the expected numbers of kinematics of the metal-poor (thick-disk/halo) population. Many of these objects were studied by \citet{2009PASP..121..117W}, who measured Fe and Ti atomic lines to demonstrate that the sdM's are metal-poor; \citet{2014A&A...568A.121N} also confirm GJ 191 and other sdM stars are metal poor. (GJ 191 is also the only M subdwarf within the the nearest one hundred star systems (www.recons.org; see \citealt{2006AJ....132.2360H}).  A third important sample is the SDSS stars with spectroscopy, although some subdwarfs were deliberately targeted and the sample includes color selections that may be biased for or against subdwarfs. \citet{2004AJ....128..426W} noted 60 SDSS M subdwarfs of all types compared to $\sim 8000$ M dwarfs.  \citet{2011AJ....141...97Wshort} identified $\sim 70,000$ spectroscopically confirmed M dwarfs in SDSS data, of which only $\sim 2000$ are M subdwarfs even including more extreme ones \citep{Bochanski:2013fk}. We take the low percentage of sdM's in these samples to support the magnitude-limited and volume-limited samples.
 \citet{Bochanski:2013fk}'s statistical parallax analysis showed that the SDSS sdM have kinematics associated with the thick disk and inner halo. \citet{2014ApJ...794..145S} have expanded the sample of the SDSS subdwarfs but reach similar conclusions.  It is notable that X-ray or UV-selected samples of M dwarfs \citep{1998ApJ...504..461F,2006AJ....132..866R,Shkolnik:2011fk} do not select many sdMs, again consistent with the old, metal-poor interpretation and not with a young, high C/O population.  Proper-motion samples such as those of \citet{2007ApJ...669.1235L} and \citet{2011AJ....142...92B} do show a much higher percentage ($\sim 20$\%) of M subdwarfs, but this is due to the bias towards the inclusion of high-velocity halo stars \citep{1975ApJ...202...22S}. To be sure, all available samples are potentially biased, and a detailed accounting of selection effects and the (unknown) relative brightness of high C/O M dwarfs compared to ordinary M dwarfs would be needed for a precise percentage, but all available evidence shows that than less than 2\% of M dwarfs could possibly be high C/O even if all sdM's are such.  Conservatively estimating at least half of sdM are metal-poor, we conclude that less than 1\% of local M dwarfs have $1.0 > $C/O$ > 0.8$.  However, we caution that if high metallicity, carbon enhanced stars exist, they would not be included in these limits.

\section{Discussion\label{discuss}}

Our model spectra show that a high carbon-to-oxygen ratio produces dramatic effects in the optical spectra of M dwarfs. While sdM stars with weak TiO relative to CaH which appear similar to the high C/O models are known in the solar neighbourhood, the vast majority of them are associated with the high-velocity, metal-poor population. M dwarfs with $1.0>$ C/O$>0.8$ and either [m/H]$<0.3$ or [m/H]$>0.3$ and [O/Fe]$<0$ are apparently rare, $<1\%$, and certainly not 10\% of the local population. However, we can place no limits on metal-rich, carbon-enhanced M dwarfs with $1.0>$ C/O$>0.8$, [m/H]$>0.3$, and [C/Fe]$>0.1$, but these objects are very rare in the G dwarf surveys.  M dwarfs and G dwarfs result from the same star formation events, so there no obvious reason why G dwarf compositions would be dramatically different than that of M dwarfs. Our results support the view \citep{Fortney:2012lr,Nissen:2013fj} that the percentage of high C/O G dwarfs was greatly overestimated.

Although not a significant population, some high C/O M dwarfs may exist, and would be of great interest for studies of stellar and exoplanet properties. Our work suggests these objects will appear as sdM's with disk-like kinematics. We cannot point to any definitive examples, but \citet{2007ApJ...669.1235L} note that some of their sdM's have disk-like kinematics, although they believe these were misclassified due to low signal-to-noise. The new large sample of SDSS sdM's \citep{2014ApJ...794..145S} with spectroscopy, including chromospheric age indicators, and kinematics may be the most promising avenue to demonstrate whether or not these stars exist. Some variation in C/O must exist, and smaller variations in C/O may contribute to scatter in the CaH/TiO indices relative to other metallicity indicators discussed by \citet{Rojas-Ayala:2012qy}. Future work should also explore the variation of other elements, such as Ti, which might also affect M dwarf spectra.

\section*{Acknowledgments}

We thank the anonymous referee for very helpful comments that improved this paper. 
JEG acknowledges the Annie Jump Cannon fund at the University of Delaware. 
PHH was supported in part by DFG grants GrK 1351 and SFB 676 C5.
The calculations presented here were performed partially
at the H\"ochstleistungs Rechenzentrum Nord (HLRN) and at the National Energy
Research Supercomputer Center (NERSC), which is supported by the Office of
Science of the U.S.  Department of Energy under Contract No. DE-AC03-76SF00098.
PHH gratefully acknowledges the Gauss Centre for Supercomputing (GCS) for providing
computing time through the John von Neumann Institute for Computing (NIC) on
the GCS share of the supercomputer JUQUEEN at J\"ulich Supercomputing Centre
(JSC). GCS is the alliance of the three national supercomputing centres HLRS
(Universit\"at Stuttgart), JSC (Forschungszentrum J\"ulich), and LRZ (Bayerische
Akademie der Wissenschaften), funded by the German Federal Ministry of
Education and Research (BMBF) and the German State Ministries for Research of
Baden-W\"urttemberg (MWK), Bayern (StMWFK) and Nordrhein-Westfalen (MIWF).

\bibliographystyle{mn2e}

\bibliography{../astrobib}
\bsp

\label{lastpage}

\end{document}